\documentclass[prb,twocolumn,superscriptaddress,citeautoscript,showpacs,amsart,longbibliography,footinbib]{revtex4}

\usepackage{graphicx}
\usepackage{multirow}
\usepackage{color}
\usepackage{bm}
\usepackage{times}
\usepackage{amsmath,bm,amsfonts}
\usepackage{dcolumn}
\usepackage{graphicx}
\usepackage{latexsym}
\usepackage{BOONDOX-cal}
\usepackage{comment}
\usepackage{appendix}
\usepackage{bibunits}

\usepackage{ulem} 
\usepackage{braket}
\usepackage{mathtools}

\usepackage{cancel}

\begin{document}

\title{Thermal Hall effect from  two-dimensional Schwinger-boson gas with Rashba spin-orbit interaction: application to ferromagnets with in-plane Dzyaloshinskii-Moriya interaction}

\author{Sungjoon \surname{Park}}

 \affiliation{Department of Physics and Astronomy, Seoul National University, Seoul 08826, Korea}

\affiliation{Center for Correlated Electron Systems, Institute for Basic Science (IBS), Seoul 08826, Korea}

\affiliation{Center for Theoretical Physics (CTP), Seoul National University, Seoul 08826, Korea}

\author{Bohm-Jung \surname{Yang}}
\email{bjyang@snu.ac.kr}
 \affiliation{Department of Physics and Astronomy, Seoul National University, Seoul 08826, Korea}

\affiliation{Center for Correlated Electron Systems, Institute for Basic Science (IBS), Seoul 08826, Korea}

\affiliation{Center for Theoretical Physics (CTP), Seoul National University, Seoul 08826, Korea}

\date{\today}

\begin{abstract}
Recently, uncovering the sources of thermal Hall effect in insulators has become an important issue.
In the case of ferromagnetic insulators, it is well known that Dzyaloshinskii-Moriya (DM) interaction can induce magnon thermal Hall effect.
Specifically, the DM vector parallel to the magnetization direction induces complex magnon hopping amplitudes, so that magnons act as if they feel Lorentz force.
However, the DM vector which is orthogonal to the magnetization direction has hitherto been neglected as a possible source of magnon thermal Hall effect.
This is because they play no role in the linear spin wave theory, an often invoked approximation when computing the magnon thermal Hall effect.
Here, we challenge this expectation by presenting the self-consistent Schwinger-boson mean field study of two-dimensional magnets with ferromagnetic Heisenberg interaction and in-plane DM interaction. 
We find that the relevant Schwinger-boson mean field Hamiltonian takes the form of two-dimensional electron gas with Rashba spin-orbit interaction, which is known to show anomalous Hall effect, spin Hall effect, and Rashba-Edelstein effect, whose thermal counterparts also appear in our system.
Importantly, the thermal Hall effect can be induced when out-of-plane magnetic field is applied, and persists even when the magnetic field is large, so that the spins are significantly polarized, and the linear spin wave theory is expected to be a reasonable approximation.
Since the linear spin wave theory predicts vanishing thermal Hall effect, our result implies that linear spin wave is not a sufficient approximation, and that magnon-magnon interaction must be taken into account to predict the correct thermal Hall conductivity.
\end{abstract}

\maketitle

\begin{bibunit}
Thermal Hall effect refers to the phenomenon in which heat current flows transversely to the temperature gradient.
When thermal Hall effect occurs in a ferromagnetic insulator, experiments \cite{onose2010observation,ideue2012effect,hirschberger2015thermal} suggest that the thermal Hall current is often dominantly carried by magnons.
In these experiments, the anomalous transport behavior of magnons was attributed to the presence of Dzyaloshinskii-Moriya (DM) interaction.
The role of the DM interaction in these experiments was understood by noticing that the component of the DM vector along the direction of the ferromagnetic order provides complex hopping amplitudes for the magnons, and thereby acts like magnetic flux to the magnons.

More generally, the origin of the intrinsic magnon thermal Hall effect was shown to be the magnon Berry curvature \cite{katsura2010theory,matsumoto2011theoretical,matsumoto2011rotational}, thus providing a topological explanation of the role played by the aforementioned DM interaction in the thermal Hall effect \cite{ideue2012effect,hirschberger2015thermal,chisnell2015topological}.
The Berry curvature formulation of the magnon thermal Hall effect was even extended to the paramagnetic regime of the ferromagnets, which was studied using the self-consistent Schwinger-boson mean field theory (SBMFT)\cite{lee2015thermal}.
It was shown that the same DM interaction responsible for the thermal Hall effect in the ferromagnetic regime also induces Berry curvature and thermal Hall effect of paramagnets  described by Schwinger-bosons in the presence of external magnetic field.

\begin{figure}[t]
\centering
\includegraphics[width=8.5cm]{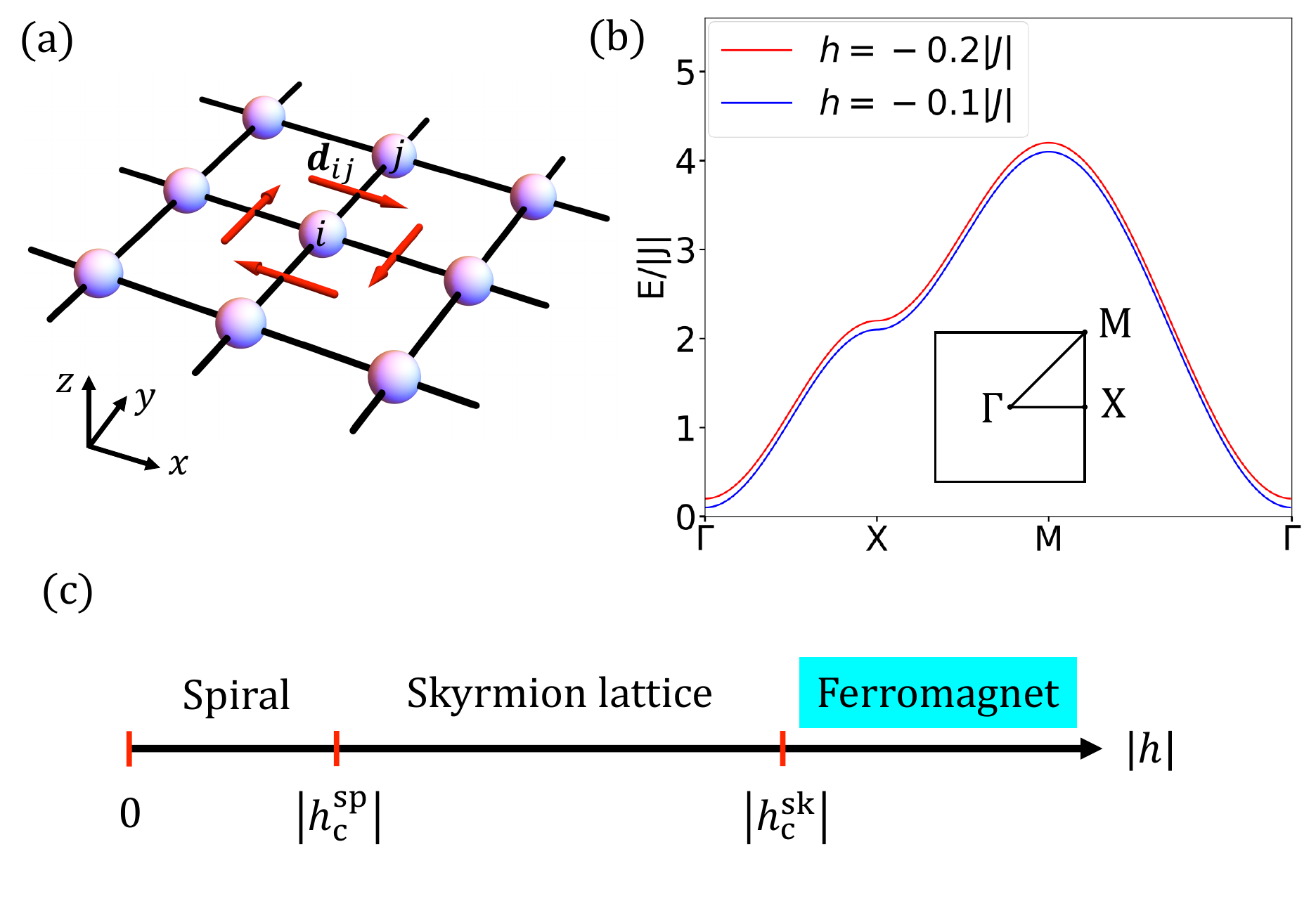}
\caption{(a) Lattice structure and DM vectors. 
(b) Magnon dispersion calculated using the linear spin wave theory with $h=-0.1 |J|$ and $h=-0.2 |J|$, where we have assumed collinear out-of-plane ferromagnetic order.
(c) The zero temperature phase diagram for our model in Eq.~\eqref{eq.model}, where the horizontal axis is the strength of the magnetic field. The critical fields are $|h|^{sp}_{c} \approx  0.0054|J|$ and $|h|^{sk}_c\approx 0.017|J|$ for our system.  Our main focus is in the ferromagnetic region.
\label{fig.lattice}}
\end{figure}

Although the DM vector parallel to the ferromagnetic order is known to produce magnon thermal Hall effect, the DM vector orthogonal to the ferromagnetic order is usually not expected to cause magnon thermal Hall effect \cite{onose2010observation,ideue2012effect}. 
For example, the  in-plane DM vectors shown in Fig.~\ref{fig.lattice}, which naturally appear when a 2D magnet is placed on a substrate, is not expected to generate magnon thermal Hall effect when the ferromagnetic order is along the $z$ direction.
This is because the DM interaction does not enter the linear spin wave theory (LSWT), which is almost always employed when computing the magnon thermal Hall effect.
However, because LSWT neglects magnon-magnon interactions, conclusions based on it may not be true, and in particular, we cannot exclude the possibility that in-plane DM interaction can contribute to thermal Hall effect when we go beyond LSWT.

In this work, we investigate the role of the in-plane DM vectors when spin fluctuation is strong by  using the self-consistent SBMFT on a minimal model, consisting of spins on a square-lattice  with nearest-neighbor ferromagnetic Heisenberg interaction and nearest-neighbor in-plane DM interaction.
We find that in the presence of in-plane DM interaction, the Schwinger-boson mean field (SBMF) Hamiltonian near the $\Gamma$ ($\bm{k}=0$) coincides with the Hamiltonian of a two-dimensional electron gas with Rashba spin-orbit coupling, so that we realize a new platform to observe the physics of Rashba system.

An important feature of our Hamiltonian is the presence of Dirac points at the time-reversal invariant momenta (TRIM), whose gap can be opened by applying an external magnetic field, resulting in Berry curvature and intrinsic thermal Hall effect.
Because the thermal Hall effect does not vanish even when there is a significant alignment of spins along the direction of magnetic field, contrary to the LSWT that predicts vanishing thermal Hall effect, our result indicates that the LSWT, which should be a good approximation when there is significant spin polarization, may not always provide a reliable approximation when computing the thermal Hall effect.
We expect that the consideration of magnon-magnon interaction can resolve the inconsistency.
Furthermore, it is well known that two-dimensional electron gas with Rashba spin-orbit coupling  shows interesting behaviors such as the Rashba-Edelstein effect \cite{bychkov1984properties,edelstein1990spin} and the  spin Hall effect \cite{sinova2004universal}.
Because our SBMF Hamiltonian also has this form, we find thermal analogues of these effects in our model.

Because the requirement for the presence of in-plane DM interaction is breaking both inversion symmetry and mirror symmetry $\mathcal{M}_z: (x,y,z) \rightarrow (x,y,-z)$, our theory is relevant to two-dimensional ferromagnets lacking  inversion and $\mathcal{M}_z$ symmetries.
These symmetries may be broken because of the intrinsic crystal structure of the magnet, or  due to the symmetry lowering arising from the substrate effect or external electric field.
Also, the in-plane DM interaction is known to stabilize skyrmion lattice, so that our theory also applies to candidate materials for skyrmion lattices.

\textit{Linear spin wave theory.|}
Let us start by presenting our minimal model and analyzing why LSWT does not predict any thermal Hall effect.
The Hamiltonian of our model is 
\begin{equation}
\mathcal{H}=\sum_{\langle i j \rangle } [J \bm{S}_i \cdot \bm{S}_j + \bm{d}_{ij}\cdot (\bm{S}_i \times \bm{S}_j)]+ \bm{h} \cdot \sum_i \bm{S}_i. \label{eq.model}
\end{equation}
Here, $J<0$ is the ferromagnetic Heisenberg interaction and $\bm{d}_{ij}$ are the nearest-neighbor DM vectors, whose directions  are indicated in Fig.~\ref{fig.lattice}. 
The magnetic field is applied along the out-of-plane direction, so that $\bm{h}=(0,0,h)$.
When the magnetic field is strong enough to polarize the spins in the $+z$ direction ($h<0$) so that we have a collinear ferromagnetic order, we can approximate the Holstein-Primakoff transformation \cite{holstein1940field} as $S^x_i\approx \frac{\sqrt{2S}}{2}(a_i+a_i^\dagger)$, $S^y_i=\frac{\sqrt{2S}}{2i}(a_i-a_i^\dagger)$, $S_i^z=S-a_i^\dagger a_i$.
Here, we have taken $\hbar=1$, and it will be restored only when necessary.
In the momentum space, we have 
\begin{equation}
\mathcal{H}_{\textrm{LSWT}}=\sum_{\bm{k}}[JS(2 \cos k_x+2 \cos k_y -4)-h]a^\dagger_{\bm{k}} a_{\bm{k}}.
\end{equation}
Notice that the DM interaction does not enter the Hamiltonian at the quadratic level.
This can easily be seen by examining the DM interaction between the $i$ and $j$ sites in Fig.~\ref{fig.lattice} (a), which is $\bm{d}_{ij}\cdot (\bm{S}_i \times \bm{S}_j)=d(S_i^yS_j^z-S_i^zS_j^y)$, where we have used $\bm{d}_{ij}=d\hat{\bm{x}}$.
By introducing the Holstein-Primakoff transformation, we see that there are no terms quadratic in the Holstein-Primakoff operators, while the linear terms cancel when we sum over the nearest neighbors.

Because there is only one magnon band, it is clear that the magnon band will not develop any Berry curvature.
This means that the LSWT does not predict any thermal Hall effect, since the intrinsic thermal Hall conductivity is given by
\begin{equation}
\kappa_{xy}=-\frac{k_B^2 T}{\hbar V} \sum_{\bm{k},n}c_2[g(E_{\bm{k},n})] \Omega_{\bm{k},n}, \label{eq.kappa_xy}
\end{equation}
where $T$ is the temperature, $V$ is the system volume, $E_{\bm{k},n}$ is the energy of $n$th magnon band, $\Omega_{\bm{k},n}$ is the Berry curvature, $g(x)$ is the Bose-Einstein distribution, and $c_2(x)=(1+x)(\ln \frac{1+x}{x})^2-(\ln x)^2-2 \textrm{Li}_2(-x)$.
Here, $\textrm{Li}_2(x)$ is the polylogarithm function $\textrm{Li}_n(x)$ for $n=2$.

Before moving on to SBMFT, let us note that at zero temperature, the Hamiltonian in Eq.~\eqref{eq.model} predicts various states other than collinear ferromagnet as shown in Fig.~\ref{fig.lattice} (c).
Adapting the mean field calculation in Ref.~[\onlinecite{ezawa2011compact}], spiral states form when the magnetic field is small, until it reaches a critical value $|h|^{sp}_{c}\approx 0.27Sd^2/|J|$. 
For $S=\frac{1}{2}$ and $d=0.2|J|$, $|h|^{sp}_c\approx 0.0054|J|$.
We note, however, that thermal Hall effect is not predicted in spiral magnets \cite{van2013magnetic}.
For magnetic field larger than $|h|^{sp}_c$, skyrmion lattice forms \cite{banerjee2014enhanced,gungordu2016stability} until it reaches another critical value  for $|h|^{sk}_c$, where $|h|^{sk}_c\approx 0.84Sd^2/|J| \approx 0.017|J|$.
For magnetic field larger than $|h|^{sk}_c$, we obtain a ferromagnet.
We note that if the skyrmion lattice state can be stabilized at finite temperature, we can expect thermal Hall conductivity to be nonzero \cite{van2013magnetic,nakata2017magnonic}.
Our main interest is the ferromagnetic regime with $|h| \gg |h|_c^{sk}$, in which case LSWT predicts  $\kappa_{xy}=0$.

\textit{Self-consistent SBMFT.|}
Recall that the Schwinger-boson representation of spin $\bm{S}_{i}$ is
\begin{equation}
\bm{S}_i=\frac{1}{2}\sum_{s,s'} b_{i,s}^\dagger \bm{\sigma}_{ss'} b_{i,s'}, \quad (s,s'=\uparrow ,~\downarrow),
\end{equation}
where $\sigma^\mu$, $\mu=x,y,z$ are the Pauli matrices. 
In this representation, the  number of the Schwinger-bosons $b_{i,s}$ ($s=\uparrow,\downarrow$) is constrained to $\mathcal{n}_i=\sum_{s} b^\dagger_{i,s}b_{i,s}=2S$ for each $i$, so that $S^z_i$ takes values in the range $-S, -S+1, ..., S$.

To carry out the SBMFT, which is reviewed in the Supplemental Materials (SM) \cite{supplement}, we define the bond operators
$\mathcal{B}_{ij}=\sum_{st}\sigma^0_{st} b_{i,s}b^\dagger_{j,t}$ and $\bm{\mathcal{C}}^\dagger_{ij}=\sum_{st}b_{i,s}^\dagger (i\bm{\sigma})_{st} b_{j,t}$, where $\sigma^0$ is the two-by-two identity matrix. 
We also adopt the notation $B=\langle \mathcal{B}_{ij} \rangle$ and $C=\frac{1}{d} \langle \bm{d}_{ij} \cdot \bm{\mathcal{C}}_{ij}\rangle$.
In the self-consistent SBMFT, we write the Hamiltonian in terms of the bond operators, carry out the Hartree-Fock decomposition, and impose the constraint $\mathcal{n}_i=2S$ on average by introducing the chemical potential $\mu$.

\begin{figure}[t]
\centering
\includegraphics[width=8.5cm]{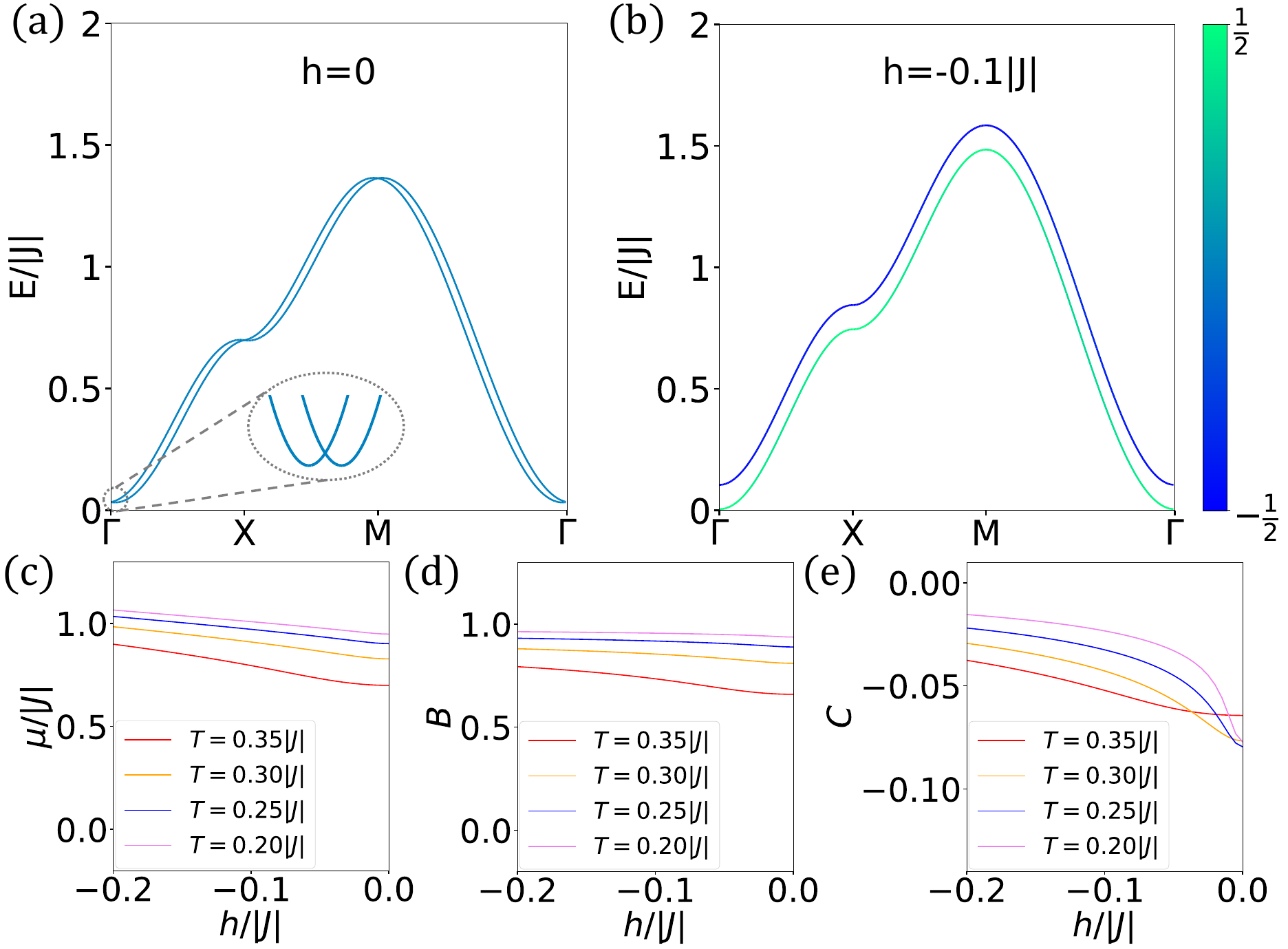}
\caption{Energy spectrum along the high symmetry lines computed using self-consistent SBMFT at $T=0.35 |J|$ when (a) $h=0$ and (b) $h=-0.1 |J|$. The color represents the expectation value of $S^z$ for the eigenstates.
The inset in (a) shows a close-up view of the Dirac point at $\Gamma$.
The solutions to the self-consistent equations are shown in (c), (d), and (e).
\label{fig.SB_ferro_hsl} }
\end{figure}

Following this procedure, the SBMF Hamiltonian is found to be 
\begin{equation}
\mathcal{H}_{\textrm{SBMF}}=\sum_{\bm{k}} \phi^\dagger_{\bm{k}} H_{\bm{k}} \phi_{\bm{k}}, \label{eq.H_tot}
\end{equation}
where $\phi_{\bm{k}} = \left(\begin{smallmatrix} b_{\bm{k},\uparrow} \\ b_{\bm{k},\downarrow} \end{smallmatrix}\right)$ and $H_{\bm{k}}=H^0_{\bm{k}}\sigma^0+\bm{h}^{\textrm{eff}}_{\bm{k}}\cdot \bm{\sigma}$, where $H^0_{\bm{k}}=\mu+(\frac{JB}{2}+\frac{Cd}{4})(\cos k_x+\cos k_y)$ and $\bm{h}^{\textrm{eff}}_{\bm{k}}=(-\frac{dB}{4}\sin k_y , \frac{dB}{4} \sin k_x, \frac{h}{2})$.
The parameters $B$, $C$, and $\mu$ in the Hamiltonian are determined by the self-consistency equations, see the SM \cite{supplement}.

We show the energy spectrum in Fig.~\ref{fig.SB_ferro_hsl} (a) and (b) for $h=0$ and $h=-0.1|J|$, respectively, using the solutions to the self-consistency equations shown in Fig.~\ref{fig.SB_ferro_hsl} (c), (d), and (e).
Notice that there is no energy degeneracy at generic momentum $\bm{k}$ even for $h=0$ because of the spin-orbit coupling provided by the in-plane DM interaction.
However, there are Dirac points at the TRIM protected by the time-reversal symmetry $\mathcal{T}$, which satisfies $\mathcal{T}^2=-1$.
In the presence of magnetic field, the broken time-reversal symmetry allows the Dirac gaps to open, and the eigenstates develop nonzero expectation value of $S^z$.
Let us note that since $\bm{h}^{\textrm{eff}}_{\bm{k}}$ can be viewed as an effective Zeeman coupling, the spin expectation value of the energy eigenstates is $\frac{1}{2} \hat{\bm{h}}^{\textrm{eff}}_{\bm{k}}$ ($-\frac{1}{2} \hat{\bm{h}}^{\textrm{eff}}_{\bm{k}}$) for the upper (lower) band, where $\hat{\bm{h}}^{\textrm{eff}}_{\bm{k}}=\frac{\bm{h}^{\textrm{eff}}_{\bm{k}}}{|\bm{h}^{\textrm{eff}}_{\bm{k}}|}$.

\begin{figure}[t]
\centering
\includegraphics[width=8.5cm]{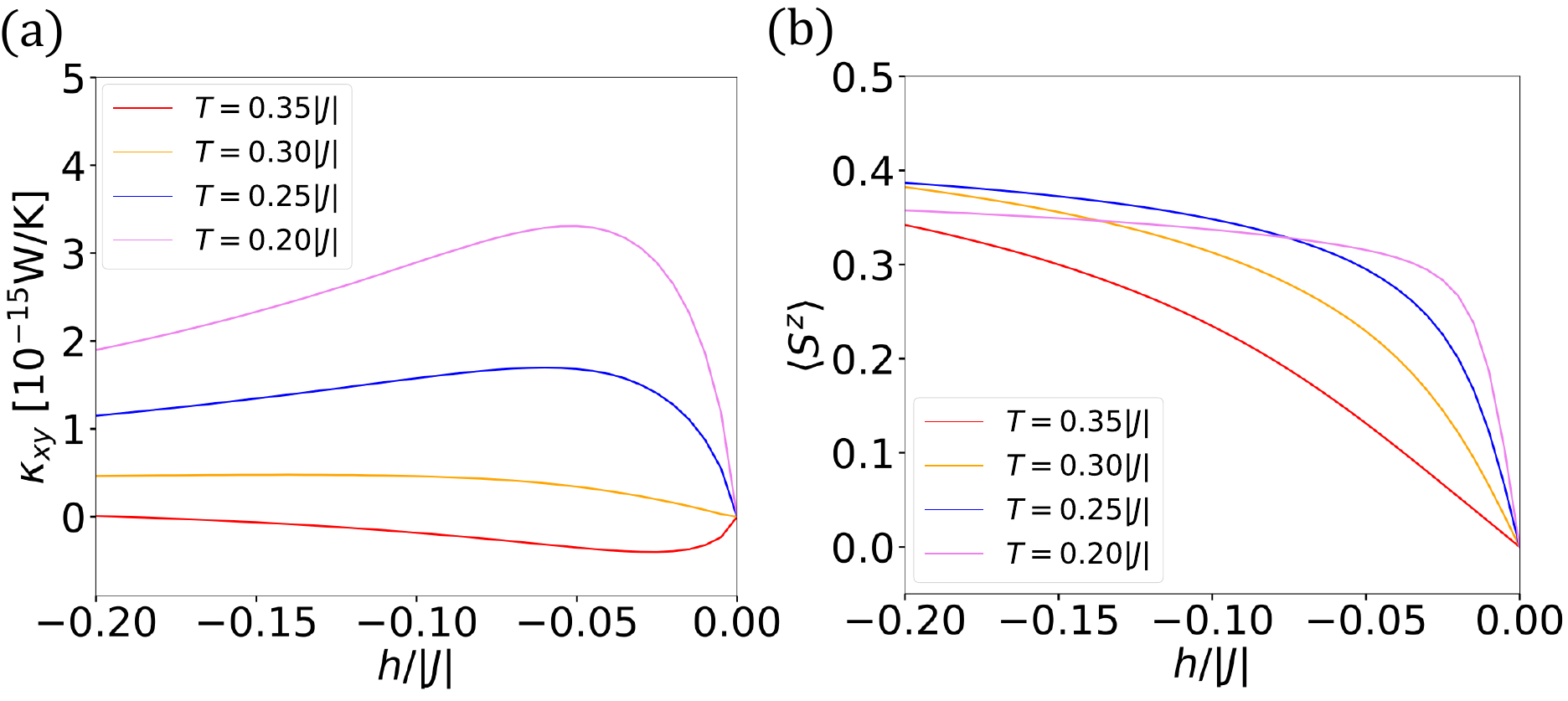}
\caption{(a) The thermal Hall conductivity and (b) the spin polarization along the $z$ direction of our 2D model in Eq.~\eqref{eq.H_tot} in the presence of out-of-plane magnetic field at various temperatures.
\label{fig.hall_and_spin}}
\end{figure}

\textit{Thermal Hall effect.|}
It is useful to compare our SBMF Hamiltonian to the spin-polarized two-dimensional electron gas with Rashba spin-orbit interaction, whose Hamiltonian is given by 
\begin{equation}
H_R=\mu_0+\frac{\bm{k}^2}{2m}+\lambda \bm{k}\cdot (\bm{\sigma}\times \hat{\bm{z}})-\Delta_0 \sigma^z, \label{eq.rashba}
\end{equation} where $\mu_0$ is the chemical potential, $m$ is the effective mass, $\lambda$ is the spin-orbit interaction strength, and $\Delta_0$ is the exchange field.
Our SBMF Hamiltonian has exactly this form near the $\Gamma$, with $\mu_0=\mu+\frac{Cd}{2}+JB$, $m=\frac{-4}{Cd+2JB}$, $\lambda=\frac{dB}{4}$, and $\Delta_0=-\frac{h}{2}$.
Notice that this Hamiltonian has a Dirac point at $\bm{k}=0$  when $\Delta_0=0$. 
When $\Delta_0 \neq 0$, the Dirac point is gapped, and the eigenstates develop non-zero Berry curvature, which results in the anomalous Hall effect in electron gas \cite{onoda2006intrinsic}.
Similarly, our SBMF Hamiltonian develops nonzero Berry curvature when magnetic field is applied, which results in nonzero thermal Hall effect.

In Fig.~\ref{fig.hall_and_spin} (a), we show the thermal Hall conductivity of our SBMF model computed using Eq.~\eqref{eq.kappa_xy}.
Notice that even when there is significant spin polarization along the $z$ axis due to the magnetic field, $\kappa_{xy \neq 0}$. 
In contrast, the LSWT, which is supposedly a good approximation when there is significant spin polarization,  predicts $\kappa_{xy}=0$.
We attribute this to the fact that linear spin wave theory does not take into account the magnon-magnon interaction.
On the other hand, because of the relations $b_{\uparrow} \leftrightarrow a$ and $b_{\downarrow} \leftrightarrow \sqrt{2S-a^\dagger a}$ between the Schwinger-bosons and the Holstein Primakoff bosons \cite{auerbach2012interacting}, magnon-magnon interaction effect is taken into account at the mean field level in the self-consistent SBMFT.

Let us note that the thermal Hall conductivity in Fig.~\ref{fig.hall_and_spin} decreases when the magnitude of the magnetic field ($|h|$) is large, especially at low temperatures. 
To understand this, we focus on the region near the $\Gamma$, where the SBMF Hamiltonian can be modeled by $H_{R}$.
The Berry curvature for $H_{R}$ is \cite{xiao2010berry} $\Omega_{\bm{k},\pm}=\mp \frac{\lambda^2 \Delta_0}{2(\lambda^2 k^2 +\Delta_0^2)^{3/2}}$, where $+$ ($-$) denotes the upper (lower) energy band.
Because the Dirac mass gap is $2|\Delta_0|=|h|$, the Dirac mass gap increases as $|h|$ increases, and as a result, the Berry curvature becomes more spread out in the $\bm{k}$ space. 
Using this, we can understand the behavior of the $\kappa_{xy}$ as a function of the magnetic field at low temperature as follows.
As we turn on the magnetic field, the Dirac gap increases, and this initially results in an increased (decreased) contribution to the thermal Hall conductivity from the lower (upper) band because the region with large Berry curvature decrease (increases) in energy, which becomes more (less) occupied due to the Bose-Einstein distribution.
For larger $|h|$, it suffices to focus on the lower band.
As $|h|$  increases, the Berry curvature spreads out in the $\bm{k}$ space. Because these states have higher energy and therefore lower occupation,  $\kappa_{xy}$ decreases.
The decrease in $\kappa_{xy}$ for large $|h|$ is consistent with the expectation that if spin wave theory with magnon-magnon interaction gives thermal hall effect, magnons will become difficult to excite thermally in the presence of large magnetic field, so that the thermal Hall effect of magnons should decrease at high magnetic field.

Finally, let us note that $\kappa_{xy}$ is an even function of $d$, which characterize the strength of the DM interaction.
To see this, note that two systems with $\bm{d}_{ij}$ and $-\bm{d}_{ij}$ are related by the $\mathcal{M}_z$ operation.
Since $\mathcal{M}_z$ changes neither the Berry curvature nor the energy, $\kappa_{xy}$ is the same for both systems.

\begin{figure}[t]
\centering
\includegraphics[width=8.5cm]{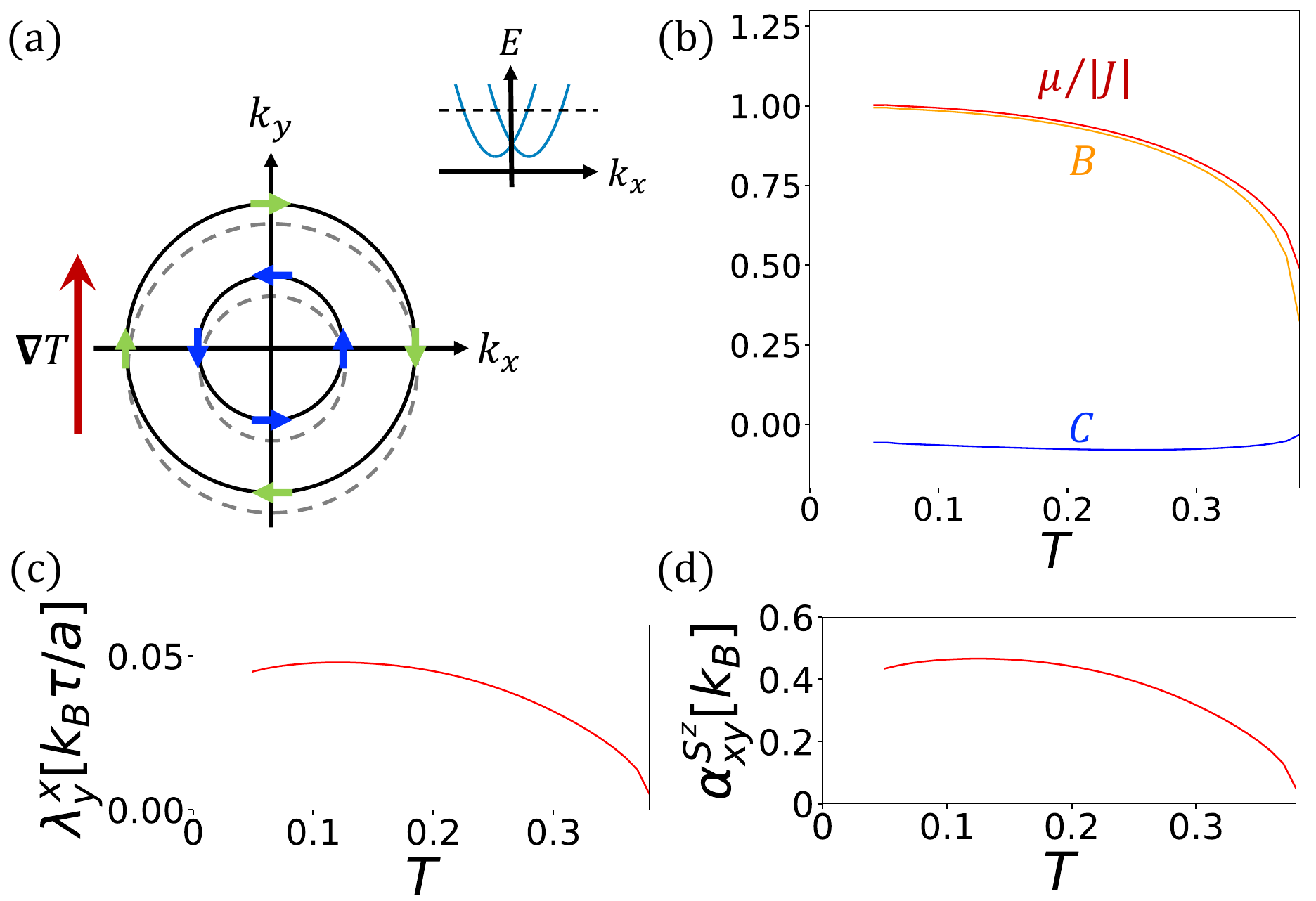}
\caption{(a) Spin-momentum locking in Eq.~\eqref{eq.rashba} shown for states at a fixed energy. 
The solid (dashed) circle indicates the states when the temperature gradient is absent (present).
(b) Temperature dependence of  solutions to the self-consistent SBMFT. 
The spin Nernst effect (c) and the thermal analogue of Rashba-Edelstein effect (d) for the solutions in (b) for the 2D model in Eq.~\eqref{eq.H_tot} in the absence of external magnetic field.
\label{fig.spin_response}}
\end{figure}

\textit{Spin response.|}
Besides the anomalous Hall effect, two-dimensional electron gas with Rashba spin-orbit interaction shows many other interesting behaviors in response to electric field, such as the Rashba-Edelstein effect \cite{bychkov1984properties,edelstein1990spin} and the  spin Hall effect \cite{sinova2004universal,inoue2004suppression,shytov2006small,sinova2006spin}.
These effects originate from the spin-momentum locking of the eigenstates of the Hamiltonian in Eq.~\eqref{eq.rashba}, which is illustrated in Fig.~\ref{fig.spin_response} (a).
We can expect thermal analogues of these effects in our model because the statistical force due to the temperature gradient can cause the spin-momentum locked states to shift in the momentum space, just as electric field does\cite{luttinger1964theory,matsumoto2011theoretical,matsumoto2011rotational}.

Let us first present the solutions to the self-consistent SBMFT as a function of temperature at zero magnetic field in Fig.~\ref{fig.spin_response} (b).
Note that there cease to be nontrivial solutions above $T_c\approx 0.381J$.  This is an artifact of the mean field approach, and it indicates that the system behaves as a paramagnet with local moments above $T_c$ \cite{arovas1988functional,yoshioka1989boson}.
Also, the self-consistent equations are difficult to solve accurately at low temperatures because of the very small Schwinger-boson energy gap. 
However, it is possible to predict the behavior of the spin responses at low temperatures analytically (see the SM \cite{supplement}).

By the thermal analogue of Rashba-Edelstein effect, we refer to the spin polarization induced by temperature gradient.
This can be estimated using the Boltzmann transport theory with constant relaxation time.
Since \cite{ashcroft1976solid} $g_{\textrm{neq}}(E)= g_{\textrm{eq}}(E)-\tau \bm{v}\cdot \bm{\nabla} T_\nu \frac{E}{k_BT^2}\frac{e^{E/k_BT}}{(e^{E/k_BT}-1)^2}$, the spin density induced by temperature gradient $(\nabla_y T) \hat{\bm{y}}$ is given by $\langle  S^\mu\rangle_{\textrm{neq}}-\langle  S^\mu \rangle_{\textrm{eq}}=-\lambda^{\mu}_{y} \nabla T_{y}$, where
\begin{align}
\lambda^\mu_{y}=&\frac{\tau}{k_BT^2}\frac{1}{V} \sum_{\bm{k},n} \langle n,\bm{k}|  S^\mu | n,\bm{k} \rangle \times\nonumber \\ 
&\langle n,\bm{k}| v_{\bm{k},y}|n,\bm{k}\rangle\frac{ E_{\bm{k},n}  e^{ E_{\bm{k},n} /k_BT}}{(e^{E_{\bm{k},n}/k_BT}-1)^2}.
\end{align}
Here, $E_{\bm{k},n}$ and $|n,\bm{k}\rangle$ are the energy and the eigenvector of $n$th energy band of $H_{\bm{k}}$, respectively. 
Due to lattice symmetries, only $\lambda^x_y$  is nonzero, which is shown in Fig.~\ref{fig.spin_response} (c) as a function of temperature.
Using the $\mathcal{M}_z$ operator as in the case of $\kappa_{xy}$, we see that $\lambda^{x}_{y}$ is an odd function of $d$.
Thus, $\lambda^{x}_{y}$ can be used to determine the sign of $d$.

Next, let us examine the spin Nernst effect, which refers to the Hall effect of spin in response to temperature gradient.
Letting $j^{S^{\mu}}_{x}$ be the current of spin $S^\mu$ along the $x$ direction, we have $j^{S^{\mu}}_{x}=-\alpha_{xy}^{S^\mu} \nabla_y T$, where the intrinsic contribution is \cite{li2020intrinsic}
\begin{equation}
\alpha^{S^\mu}_{xy}=\frac{ 2k_{B}\hbar}{V}\sum_{\bm{k},n} (\Omega^{S^\mu}_{\bm{k},n})_{xy} c_1(E_{\bm{k},n}),
\end{equation}
$(\Omega^{S^\mu}_{\bm{k},n})_{xy} = \sum_{m}' \frac{\textrm{Im}[\langle n,\bm{k}| S^z v_{\bm{k},x}+v_{\bm{k},x} S^z | m, \bm{k} \rangle \langle m, \bm{k}| v_{\bm{k},y}| n,\bm{k} \rangle ]}{(E_{\bm{k},n}-E_{\bm{k},m})^2} $, where$~'$ indicates that the sum excludes $m=n$, $v_{\bm{k},\mu}=\frac{1}{\hbar} \frac{\partial H_{\bm{k}}}{\partial k_\mu}$, and $c_1(x)=(1+g(x)) \log(1+g(x))-g(x)\log g(x)$.
Due to lattice symmetries,  $\alpha^{S^\mu}_{xy}\neq 0$ only for $\mu=z$, which we show in Fig.~\ref{fig.spin_response} (d).
Using the $\mathcal{M}_z$ operator as before, we find that $\alpha^{S^z}_{xy}$ is an even function of $d$.

\textit{Discussion.|}
In this work, we have investigated thermal transport signatures of in-plane DM interaction in a two-dimensional paramagnet with ferromagnetic exchange interaction.
Although the common expectation is that in-plane DM interaction is not important for thermal Hall effect when spins are aligned in the out-of-plane direction, we have shown that this is not the case by using the self-consistent SBMFT.
Since the LSWT predicts no thermal Hall effect, our result implies that magnon-magnon interaction must be considered to predict even a qualitatively correct magnon thermal Hall effect.
It is also important to note that the magnitude of the thermal Hall conductivity is in experimentally measurable range.
To see this, let us stack our paramagnetic model in Eq.~\eqref{eq.model} along the $z$ axis with interlayer distance of a few angstroms.
Then, the 3D thermal Hall conductivity can be expected to be around $10^{-5}$W/Km, which is comparable to the phonon thermal Hall effect measured in paramagnets \cite{strohm2005phenomenological,mori2014origin}.
Finally, we have also shown that the in-plane DM interaction can induce spin Nernst effect as well as thermal version of Rashba-Edelstein effect.

Let us note that the DM interaction we have studied is the type that can stabilize the Neel type skyrmion.
However, our theory can trivially be extended to include the DM interaction which can stabilize Bloch type skyrmion, and which is often written in the form $D \bm{n}\cdot  (\bm{\nabla} \times \bm{n}) $, as they are related by global rotation of spins about the $z$-axis.
Thus, our theory is applicable to two-dimensional ferromagnets which are Bloch and Neel type skyrmion candidates.
As examples, we propose $\textrm{CrI}_3$ and $\textrm{Cr}_2\textrm{Ge}_2\textrm{Te}_6$, which are well known two-dimensional ferromagnets \cite{huang2017layer,gong2017discovery}.
In both cases, there is an inversion symmetry between the nearest-neighbor Cr atoms, so that DM interaction is forbidden.
However, when the magnets are placed on a substrate or when out-of-plane electric field is applied \cite{behera2019magnetic}, the inversion symmetry is broken, and the in-plane DM interaction can be induced, and we expect our theory to apply in this setup.

\begin{acknowledgements}
S.P. was supported by IBS-R009-D1. B.-J.Y. was supported by the Institute for Basic Science in Korea (Grant No. IBS-R009-D1) and Basic Science Research Program through the National Research Foundation of Korea (NRF) (Grant No. 0426-20200003). This work was supported in part by the U.S. Army Research Office under Grant Number W911NF-18-1-0137.
\end{acknowledgements}

\end{bibunit}

\clearpage

%
\onecolumngrid
\begin{center}
\large{\textbf{Supplemental Material for \\``Thermal Hall effect from  
two-dimensional Schwinger-boson gas with Rashba spin-orbit interaction: 
application to ferromagnets with in-plane Dzyaloshinskii-Moriya interaction"}} \\~\\
\normalsize{Sungjoon Park and Bohm-Jung Yang} \\~
\end{center}
\twocolumngrid

\setcounter{equation}{0}
\setcounter{figure}{0}
\setcounter{page}{1}
\renewcommand{\theequation}{{S}\arabic{equation}}
\renewcommand{\thefigure}{{S}\arabic{figure}}
\renewcommand{\thesection}{{SM~}\arabic{section}}

\begin{bibunit}
\section{Review of SBMFT}
Let us define the bond operators
\begin{align}
\mathcal{A}_{ij}&=\sum_{st}\epsilon_{st} b_{i,s} b_{j,t}, \nonumber \\
\mathcal{B}_{ij}&=\sum_{st}\sigma^0_{st} b_{i,s}b^\dagger_{j,t}, \nonumber \\
\bm{\mathcal{C}}^\dagger_{ij}&=\sum_{st}b_{i,s}^\dagger (i\bm{\sigma})_{st} b_{j,t}, \nonumber \\
\bm{\mathcal{D}}_{ij}&=\sum_{st}b_{i,s}(\sigma^y \bm{\sigma})_{st} b_{j,t}.\label{eq.bond_operators}
\end{align}
To write the Heisenberg and DM interactions in terms of the bond operators, it is useful to note the identities
\begin{align}
\bm{\sigma}_{ss'}\cdot \bm{\sigma}_{tt'}&=\sigma^0_{ss'}\sigma^0_{tt'}-2\epsilon_{st}\epsilon_{s't'} \nonumber \\
&=-\sigma^0_{ss'} \sigma^0_{tt'}+2\sigma^0_{st'}\sigma^0_{s't} \label{eq.pauli_identity1}, \\
\bm{\sigma}_{ss'} \times \bm{\sigma}_{tt'}&=\begin{bmatrix}
(\sigma^x\sigma^y)_{st} \epsilon_{s't'}+\epsilon_{st} (\sigma^y\sigma^x)_{s't'}  \nonumber \\
(\sigma^y\sigma^y)_{st}\epsilon_{s't'}+\epsilon_{st}(\sigma^y\sigma^y)_{s't'}  \\
(\sigma^z \sigma^y)_{st}\epsilon_{s't'}+\epsilon_{st}(\sigma^y\sigma^z)_{s't'} 
\end{bmatrix}\\
&=\begin{bmatrix}
i\sigma^x_{st'} \sigma^0_{s't}-i\sigma^0_{st'} \sigma^x_{s't} \\
i\sigma^y_{st'}\sigma^0_{s't}+i\sigma^0_{st'}\sigma^y_{s't} \\
i\sigma^z_{st'} \sigma^0_{s't}-i\sigma^0_{st'} \sigma^z_{s't} 
\end{bmatrix}, \label{eq.pauli_identity2}
\end{align}
where $\sigma^0$ is the two by two identity matrix and $\epsilon=i\sigma^y$.

Using Eq.~\eqref{eq.pauli_identity1}, we have
\begin{align}
\bm{S}_i\cdot \bm{S}_j&=\frac{1}{2}[2S^2-\mathcal{A}_{ij}^\dagger \mathcal{A}_{ij}] \nonumber \\
&=\frac{1}{2}[\mathcal{B}_{ij}^\dagger \mathcal{B}_{ij} -2S(S+1)],\label{eq.heisenberg_bond}
\end{align}
and using Eq.~\eqref{eq.pauli_identity2}, we have
\begin{align}
\bm{S}_i \times \bm{S}_j=\frac{1}{8}[\mathcal{A}^\dagger_{ij} \bm{\mathcal{D}}_{ij}+\bm{\mathcal{D}}^\dagger_{ij}\mathcal{A}_{ij}+:\mathcal{B}^\dagger_{ij}\bm{\mathcal{C}}_{ij}:+:\bm{\mathcal{C}}_{ij}^\dagger \mathcal{B}_{ij}:],
\end{align}
where $:~:$ indicates normal ordering.
From the identities
\begin{align}
:\bm{\mathcal{C}}_{ij}^\dagger \mathcal{B}_{ij}:&=\bm{\mathcal{C}}^\dagger_{ij} \mathcal{B}_{ij}-2i\bm{S}_{i}\\
:\mathcal{B}^\dagger_{ij}\bm{\mathcal{C}}_{ij} :&=\mathcal{B}^\dagger_{ij} \bm{\mathcal{C}}_{ij}+2i\bm{S}_{i},
\end{align}
it follows that
\begin{align}
\bm{S}_i \times \bm{S}_j=\frac{1}{8}[\mathcal{A}^\dagger_{ij} \bm{\mathcal{D}}_{ij}+\bm{\mathcal{D}}^\dagger_{ij}\mathcal{A}_{ij}+\mathcal{B}^\dagger_{ij}\bm{\mathcal{C}}_{ij}+\bm{\mathcal{C}}_{ij}^\dagger \mathcal{B}_{ij}].\label{eq.dm_bond}
\end{align}
Note that $\mathcal{A}_{ij}$ and  $\mathcal{B}_{ij}$ are called antiferromagnetic and ferromagnetic bond operators, respectively, because for ferromagnetic Heisenberg interaction ($J<0$), it is energetically favorable to have ferromagnetic bond over antiferromagnetic bond, while the opposite is true for antiferromagnetic Heisenberg interaction ($J>0$). 

To carry out the mean field theory, we use the identity
\begin{equation}
\mathcal{O}_1 \mathcal{O}_2=\delta \mathcal{O}_1\delta \mathcal{O}_2+\mathcal{O}_1 O_2 +\mathcal{O}_2 O_1 -O_1 O_2,
\end{equation}
where $\delta{\mathcal{O}}_i=\mathcal{O}_i-\langle \mathcal{O}_i \rangle$ and $O_i=\langle \mathcal{O}_i \rangle$.
The Hartree-Fock decomposition of the $J\bm{S}_i \cdot \bm{S}_j$ is
\begin{align}
J\bm{S}_i \cdot \bm{S}_j=E_{ij}^{J_0}+E_{ij}^{J_{\textrm{int}}}+E_{ij}^{J_{\textrm{free}}},
\end{align}
where $E_{ij}^{J_0}=\frac{JS}{2}+\frac{J}{4} [A_{ij}^* A_{ij}-B_{ij}^* B_{ij}]$ is the zero point energy and $E_{ij}^{J_{\textrm{int}}}=\frac{J}{4}[-\delta A_{ij}^\dagger \delta A_{ij}+\delta B_{ij}^\dagger \delta B_{ij}] $ is the energy from fluctuation about the mean field solution, which contains the interaction terms between Schwinger-bosons.
For mean field theory, we focus on the part quadratic in the Schwinger-bosons, which is given by
\begin{equation}
E_{ij}^{J_{\textrm{free}}}=\frac{J}{4} [B_{ij}^* \mathcal{B}_{ij}+\mathcal{B}_{ij}^\dagger B_{ij}-A_{ij}^* \mathcal{A}_{ij}-\mathcal{A}_{ij}^\dagger A_{ij}].\label{eq.Heisenberg_SB}
\end{equation}
The DM interaction can be similarly decomposed, with 
\begin{align}
E_{ij}^{D_{\textrm{free}}}=&\frac{\bm{d}_{ij}}{8}\cdot[A_{ij}^* \bm{\mathcal{D}}_{ij}+\mathcal{A}^\dagger_{ij}\bm{D}_{ij}
+\bm{D}_{ij}^*\mathcal{A}_{ij} +  \bm{\mathcal{D}}^\dagger_{ij}A_{ij}\nonumber \\
&+B^*_{ij}\bm{\mathcal{C}}_{ij}+\mathcal{B}^\dagger_{ij}\bm{C}_{ij}
+\bm{C}_{ij}^* \mathcal{B}_{ij}+\bm{\mathcal{C}}_{ij}^\dagger B_{ij}]. \label{eq.DM_SB}
\end{align}

Now, let us consider the model in Eq.~\eqref{eq.model} using Eqs.~\eqref{eq.Heisenberg_SB} and \eqref{eq.DM_SB}.
Using the $U(1)$ gauge symmetry $b_{i,s}\rightarrow e^{i\phi_{i}}b_{i,s}$ of the Schwinger-bosons, we can fix the gauge so that $B\equiv \langle \mathcal{B}_{ij} \rangle=|B|$.
We assume that the solution to the SBMFT does not break the translation symmetry, the four-fold rotation symmetry, and the mirror symmetries about the planes normal to the $x$ and the $y$ axes.
Note also that because $J<0$, antiferromagnetic bond operators are disfavored. 
We can thus assume that there are no anomalous terms (i.e. terms that does not conserve the number of Schwinger-boson number) in the SBMF Hamiltonian. 
Note that this expectation is not changed by the presence of a small DM interaction.
To see this, note that classically, the energy due to the anomalous terms between a pair of spins is $-\frac{J}{4} |A_{ij}|^2 +\frac{d}{4} \textrm{Re} (A^*_{ij} D_{ij})$. 
Since $A_{ij}=0$ and $D_{ij}=0$ for $d=0$, $A_{ij}=\alpha d+O(d^2)$ and $D_{ij}=\delta d+O(d^2)$ for some constants $\alpha$ and $\delta$. 
This implies that for small $d$, it is energetically unfavorable to have nonzero $A_{ij}$ or $D_{ij}$.
Therefore, we restrict to ansatz with no anomalous terms in this work.
We further note that by imposing the fourfold rotation symmetry and the translational symmetry, we have $C\equiv \frac{1}{d} \langle \bm{d}_{ij} \cdot \bm{\mathcal{C}}_{ij}\rangle = \textrm{Re}(C)$.

We can now obtain the SBMF Hamiltonian $\mathcal{H}_{\textrm{SBMF}}$ in the main text is obtained as follows.
Fourier transforming the nearest-neighbor ferromagnetic Heisenberg interaction that is quadratic in the Schwinger-bosons, we obtain
\begin{align}
\mathcal{H}_J&=\frac{J B}{2}\sum_{\bm{k}} \phi_{\bm{k}}^\dagger \sigma^0 \phi_{\bm{k}}[\cos k_x + \cos k_y],
\end{align}
where $\phi_{\bm{k}} = \left(\begin{smallmatrix} b_{\bm{k},\uparrow} \\ b_{\bm{k},\downarrow} \end{smallmatrix}\right)$.
Similarly, the nearest-neighbor DM interaction becomes
\begin{align}
\mathcal{H}_{D}=&\frac{dB}{4}\sum_{\bm{k}} \phi^\dagger_{\bm{k}} (\sin k_x \sigma^y-\sin k_y \sigma^x) \phi_{\bm{k}} \nonumber \\
&+\frac{Cd}{4}\sum_{\bm{k}} \phi^\dagger_{\bm{k}} \sigma^0 (\cos k_x + \cos k_y) \phi_{\bm{k}}.
\end{align}
The Zeeman interaction due to the out-of-plane magnetic field is
\begin{equation}
\mathcal{H}_h=\frac{h}{2} \sum_{\bm{k}} \phi^\dagger_{\bm{k}} \sigma^z \phi_{\bm{k}}.
\end{equation}
We must also include the chemical potential term $\mu \sum_{\bm{k}}(\phi^{\dagger}_{\bm{k}} \sigma^0 \phi_{\bm{k}}-2S)$ to enforce the constraint $\langle \mathcal{n}_i \rangle = \sum_{s} \langle b^\dagger_{i,s}b_{i,s}\rangle=2S$ on average. 
Let
\begin{equation}
\mathcal{H}_\mu=\mu \sum_{\bm{k}}b^{\dagger}_{\bm{k}} \sigma^0 b_{\bm{k}}.
\end{equation}
be the quadratic part of the chemical potential term.
The SBMF Hamiltonian is given by
\begin{equation}
\mathcal{H}_{\textrm{SBMF}}=\mathcal{H}_J+\mathcal{H}_D+\mathcal{H}_h+\mathcal{H}_\mu = \sum_{\bm{k}} \phi^\dagger_{\bm{k}} H_{\bm{k}} \phi_{\bm{k}},
\end{equation}
which is Eq.~\eqref{eq.H_tot} in the main text.

Finally, let us note that by denoting the energy and the eigenvector of $n$th energy band of $H_{\bm{k}}$ by $E_{\bm{k},n}$ and $|n,\bm{k}\rangle$, respectively, the self-consistency equations are 
\begin{align}
& 2S = \frac{1}{N}\sum_{\bm{k},n}g(E_{\bm{k},n}), \\ 
& B=\frac{1}{2N} \sum_{\bm{k},n} g(E_{\bm{k},n})(\cos k_x+\cos k_y), \\
& C=\frac{1}{2N} \sum_{\bm{k},n}g(E_{\bm{k},n}) \langle n, \bm{k}| [\sigma^y \sin k_x - \sigma^x \sin k_y]|n,\bm{k}\rangle.
\end{align}
These equations can be solved by using the Levenberg-Marquardt method of least squares\cite{misguich2012schwinger} to minimize  $(\langle \mathcal{n}_i \rangle-2S)^2+(\langle \mathcal{B}_{ij}\rangle-B)^2+(\langle \mathcal{C}_{ij}\cdot \bm{d}_{ij}\rangle/d - C )^2$.

\section{Low temperature behavior of $\alpha^{S^z}_{xy}$ and $\lambda^x_y$}  \label{sec.t_behavior}
Let us find the low temperature behavior of the conductivities.
The energy is $E_{\bm{k}, \pm}=a(\cos k_x + \cos k_y)+\mu \pm \sqrt{b^2(\sin^2 k_x + \sin^2 k_y)}$, where $a=\frac{2JB+Cd}{4}$ and $b=\frac{dB}{4}$.
It is useful to note that $a, b, \mu = \textrm{const}.+O(T)$.
When $d$ is small, we can expand the Hamiltonian by assuming that $k_x$ and $k_y$ are small to obtain the two-dimensional Rashba Hamiltonian, see the main text. 
Then, the locus of $\bm{k}$ with minimum energy forms a circle with $k=m\lambda$ and $E_{-,k=m\lambda}=\mu-\frac{m\lambda^2}{2}$, and the Schwinger-boson energy gap is proportional to $T^2$ at low temperature \cite{auerbach2012interacting}. 
Note that this implies that the phase transition to ordered phase (condensation of Schwinger-bosons) occurs at zero temperature (because we have not included magnetic anisotropy) with finite ordering vector ($k=m\lambda$).

Straightforward calculation shows that for two-dimensional Rashba gas,
\begin{align}
\hbar (\Omega_{\bm{k},\pm}^{S^z})_{xy}=\mp \frac{k_x^2}{4\lambda k^3}.
\end{align}
Thus, $(\Omega_{\bm{k},\pm}^{S^z})_{xy} \propto \cos ^2 \theta$ near $k=m\lambda$ $(\cos \theta = \frac{k_x}{k})$.
Then, $\alpha^{S^z}_{xy}\propto \int kdk c_1(E_{-,\bm{k}})=\frac{\sqrt{T}}{2}\int dx \frac{m\lambda+\sqrt{Tx}}{\sqrt{x}}[\alpha T+\beta x-\log(e^{\alpha T+\beta x}-1) + \frac{\alpha T+\beta x}{e^{\alpha T+\beta x}-1}]$, where $x=(k-m\lambda)^2/T$ and $E_{-}=\alpha T^2+\beta (k-m\lambda )^2$.
Because the integral converges to a constant as $T\rightarrow 0$, $\alpha^{S^z}_{xy}\propto \sqrt{T}$ to the lowest order in $T$.

We can similarly analyze the temperature dependence of $\lambda^x_y$.
Using $\langle -,\bm{k}| S^x | -, \bm{k} \rangle=\frac{S}{2} \sin \theta$, 
$\langle -,\bm{k}| v_y | -, \bm{k} \rangle=\frac{\sin \theta (k-\lambda m)}{m}$, we find that the leading order in $T$ is
 $\lambda^x_y \propto \frac{1}{\sqrt{T}} \int dx   \frac{x^2 (\alpha T^2 + T\beta x^2) e^{\alpha T+ \beta x^2}}{(e^{\alpha T+ \beta x^2}-1)^2}$, where $k=\sqrt{T} x+ \lambda m$.
From this, we find that the leading order $T$ is given by $\lambda^x_y\propto \sqrt{T}$.


\end{bibunit}
\end{document}